\documentclass[a4paper, amsfonts, amssymb, amsmath, reprint, showkeys, nofootinbib]{revtex4-1}
\usepackage[english]{babel}
\usepackage[utf8]{inputenc}
\usepackage[colorinlistoftodos, color=green!40, prependcaption]{todonotes}
\usepackage[pdftex, pdftitle={Article}, pdfauthor={Author}]{hyperref}
\usepackage{float}

\usepackage{graphicx}
\usepackage{yhmath}
\usepackage{subfigure}
\usepackage{mathdots}
\usepackage{MnSymbol}
\newcommand{\nb}[1]{\textcolor{black}{#1}}
\newcommand{\ak}[1]{\textcolor{black}{#1}}
\begin{document}
\title{Coupling Light with Matter for Identifying Dominant Subnetworks}
	\author{Airat Kamaletdinov and Natalia G. Berloff }
	\email[correspondence address: ]{N.G.Berloff@damtp.cam.ac.uk}	
	\affiliation{Department of Applied Mathematics and Theoretical Physics, University of Cambridge, Cambridge CB3 0WA, United Kingdom}

\date{\today}

\begin{abstract}
\nb{We introduce DOMINO, a light–matter computing platform that exploits the full complex amplitude of coupled condensate networks to solve maximum-weight clique problems and reveal hidden indirect correlations in large graphs. By embedding network structure directly into a gain-controlled polaritonic (or photonic) oscillator array, DOMINO performs analog optimization, directly solving the maximum‑weight clique problem via the gain–controlled minimisation, through a physically enforced global-intensity constraint, allowing the system to converge rapidly to dominant subnetworks while simultaneously extracting phase, encoded co- and counter-regulation patterns. This gain-based mechanism unlocks capabilities inaccessible to conventional Ising-type simulators: all degrees of freedom (amplitude and phase) participate in the computation, dramatically expanding the class of problems that can be efficiently encoded. Our approach is inherently ultrafast, energy-efficient, and naturally robust to noise, requiring no digital post-processing. Applied to real gene–gene coexpression data, DOMINO reliably identifies biologically meaningful transcription-regulator modules and exposes latent regulatory relationships. Because the method applies generically to any weighted network, it establishes a scalable physical route to solving high-value graph-analytic tasks across biology, finance, social systems, and engineered networks.}

\end{abstract}

\maketitle
\section{Introduction}

 \nb{Recent years have seen rapid progress in specialised computing architectures
designed to accelerate optimisation and machine-learning workloads by exploiting
physical dynamics directly. These systems, often described as physical neural
networks (PNNs), operate by steering a physical medium toward low-loss or
low-energy states using principles such as thermal and quantum annealing, Hopf
bifurcation at condensation thresholds, minimum-power dissipation, the principle
of least action, and minimum-entropy production \cite{vadlamani2020physics}.  These
principles underpin the broader framework of physics-inspired heuristics known as
$\pi$-computing \cite{cummins2025vector,cummins2025ising}.
Hardware demonstrations already highlight the potential of this approach: the
Microsoft Analog Optical Computer (AOC) \cite{kalinin2023analog,kalinin2025analog}
achieves state-of-the-art performance on hard quadratic-programming instances
(QPLIB), with up to three orders of magnitude shorter time-to-solution than
Gurobi and projected $\sim 100\times$ energy-efficiency gains over leading GPUs.
Coherent Ising machines based on degenerate optical parametric oscillators solve
MAX-CUT and Sherrington–Kirkpatrick problems with problem-size–independent
runtime and significantly outperform sparsely connected quantum annealers on
comparable benchmarks \cite{haribara2016coherent,haribara2017performance,hamerly2019experimental}.
Related ideas have been translated into efficient CMOS/FPGA/GPU algorithms such
as the Digital Annealer \cite{aramon2019physics} and simulated bifurcation
methods \cite{goto2019combinatorial}, which achieve order-of-magnitude
time-to-solution improvements over classical simulated annealing.
Within this landscape, coupled light–matter complex-valued neural networks offer
a particularly powerful class of PNNs. Exciton–polariton condensates epitomise
the light–matter coupling behind gain-based computing (GBC): a regime in which
gain control, nonlinear interactions, and phase coherence jointly drive the system
toward loss-minimising states. GBC leverages modern laser technologies and spatial
light modulators to realise massively parallel, highly nonlinear analogue
computation.
A key advantage of these platforms is the ability to exploit the \emph{full}
complex degree of freedom of light. Recent analyses demonstrate that
complex-valued optical and polaritonic architectures can encode up to t
}

\nb{In contrast to phase-only or binarised optical Ising machines, such
complex-valued neurons provide both amplitude and phase degrees of freedom,
which can be harnessed to encode weighted connectivity and correlation
structure simultaneously. As argued in recent analyses of all-optical neural
networks~\cite{opticalNNs}, exploiting complex-valued signals can
significantly increase the expressive power and efficiency of photonic
architectures; DOMINO leverages exactly this advantage by using amplitudes for
combinatorial structure and phases for correlations within the same physical
platform.}

Recent advancements in GBC-inspired hardware have introduced a variety of technologies. These include coherent Ising machines (CIMs) based on optical parametric oscillators (OPOs) \cite{mcmahon2016fully, inagaki2016coherent, yamamoto2017coherent, honjo2021100},  advanced laser systems \cite{babaeian2019single, pal2020rapid, parto2020realizing}, photonic simulators \cite{pierangeli2019large, roques2020heuristic},  polariton \cite{berloff2017realizing, kalinin2020polaritonic} and photon condensates \cite{vretenar2021controllable}.
 The primary focus of these approaches is to solve optimization problems on a set of either discrete or continuous variables that can be mapped onto the phases of generally complex signal amplitudes so that the goal is limited to finding the ground state of various classical spin Hamiltonians such as  Ising, XY, Clock, k-local, etc. \cite{lucas2014ising}. \nb{Within this family, exciton--polariton condensates~\cite{berloff2017realizing,kalinin2018networks,kalinin2019polaritonic}
have emerged as a particularly promising neuromorphic platform, demonstrating
high-speed analog optimization and pattern recognition in lattice geometries.
DOMINO builds directly on this polaritonic hardware paradigm, but repurposes it
from emulating spin models to solving maximum-weight clique and dominant
subnetwork problems on arbitrary weighted graphs. Among algorithmic developments in this class, the Vector Ising Spin Annealer (VISA) extends
conventional Ising-based heuristics by relaxing binary spins into continuous vector degrees of 
freedom, enabling smoother annealing trajectories and improved exploration of the energy 
landscape \cite{cummins2025vector}. However, VISA remains restricted to spin-Hamiltonian ground states, whereas DOMINO leverages complex amplitudes and a global-intensity constraint to solve a non-spin quadratic programme  directly.
}

However, supporting only the problems that can be mapped into a standard spin Hamiltonian restricts the scope of applications that can be efficiently embedded in the physical hardware or algorithms. Notably, there have been several proposals to extend the class of Ising Hamiltonians to mixed-integer or box-constrained quadratic optimization problems \cite{khosravi2021mixed,khosravi2022non,kalinin2023analog}. 

 \nb{Our approach fits naturally into this gain-based computing (GBC) landscape, but
goes beyond the usual focus on classical spin Hamiltonians. Whereas most GBC
implementations target Ising-, XY-, or clock-type models with binary or
phase-only degrees of freedom, DOMINO uses gain-driven complex oscillators to
solve a graph-theoretic optimization problem directly: the maximum-weight
clique problem on weighted networks. In this sense, DOMINO extends the scope of
GBC from spin Hamiltonians to dominant-subnetwork detection, while retaining
the same physical ingredients of gain, saturation, and coherent coupling.
}

In our paper, we show that the amplitudes and phases of complex-valued oscillators can be exploited to solve the problem of identifying a dominant subnetwork—a problem central to network analysis. \nb{More specifically, we (i) introduce DOMINO, a gain-based optical solver
for the maximum-weight clique problem that exploits
both amplitudes and phases of complex-valued oscillators;
(ii) demonstrate on both structured and random
weighted networks that DOMINO achieves higher success
probabilities and shorter times-to-solution than a representative
quasi--Newton method (BFGS) on the continuous Motzkin--Straus
objective, and is competitive with a state-of-the-art commercial
solver (Gurobi) on the benchmarks considered; and
(iii) apply DOMINO to a real transcription-regulator coexpression
network, extracting biologically meaningful dominant cliques
and illustrating how the same hardware can infer indirect correlations
via phase dynamics.
 }

 \textit{Maximum weighted clique problem.—} Identifying subnetworks of closely interacting nodes
can provide valuable insight into the structure and function of complex systems. For example, in social networks,
identifying sub-communities can help understand how information spreads within the network and how different
groups interact. In financial networks, identifying clusters of companies or markets can help understand how
different sectors of the economy are interconnected and how they influence each other. In transportation networks,
identifying clusters of locations can help optimise transportation routes and improve efficiency. Identifying
groups of devices or users in communication networks can help improve network performance and reliability.
The analysis of subnetworks within complex biomolecular networks offers valuable insights into the intricate patterns
of cellular interactions, particularly in the context of temporal and condition-specific conditions. Extracting
condition-specific subnetworks is crucial in understanding cellular adaptations to environmental changes
and gene expression alterations associated with diseases and ageing.
\ak{Biomolecular networks are graph-based representations of how genes are regulated in coordination. Each node
represents a gene, and edges indicate how similarly pairs of genes are activated across various biological
conditions. This similarity, known as co-expression, is often quantified using correlation-based matrices
$\mathbf{J}$, with components $J_{ij}$ that characterise the strength and kind (positive or negative) of the
correlations between the $i$-th and $j$-th genes.}

\nb{From a graph-theoretic perspective, such a network can be represented as a weighted undirected graph
$G=(V,E,w)$, where $V$ is the set of vertices (genes), $E\subseteq V\times V$ is the set of edges, and
$w:E\to\mathbb{R}_{\ge0}$ assigns a nonnegative weight $w_{ij}$ to each edge $(i,j)\in E$ (for example,
$w_{ij} = |J_{ij}|$). A subset $C\subseteq V$ is a \emph{clique} if every pair of distinct vertices in $C$ is
connected by an edge. The weight of a clique $C$ is then
\[
W(C) = \sum_{\substack{i<j\\ i,j\in C}} w_{ij}.
\]
The \emph{maximum–weight clique problem} (MWCP) is the task of finding a clique
\[
C^\star \in \arg\max_{C\subseteq V\ \text{clique}} W(C),
\]
i.e. a fully connected subnetwork whose total edge weight is maximal among all cliques in the graph
\cite{amgalan2014wmaxc}. Throughout this work we use maximum–weight clique (MWC) and dominant subnetwork
interchangeably; in the biological setting, the MWC corresponds to a tightly coexpressed gene module and plays a
crucial role in understanding biological functions (pathways) of simultaneously co-expressed genes.}

\nb{In conventional in~silico analyses, such dominant or densely connected
subnetworks are typically identified by weighted gene coexpression network
analysis and related module-detection methods, which infer coexpressed
modules numerically from large transcriptomic datasets \cite{langfelder2008wgcna}. Our optical DOMINO
framework offers a complementary route: a hardware accelerator that can
identify such dominant subnetworks directly in analog form, with amplitudes
encoding clique membership and weights.}

There are different ways to \ak{find} the dominant subnetwork given the coupling matrix $\mathbf{J}$. One common
approach is to find the principal eigenvector of $\mathbf{J}$, which gives the steady-state distribution of a
Markov chain in a random-walk model where transition probabilities are proportional to the edge weights. Nodes
with large components in the dominant eigenvector are then interpreted as strongly connected or influential.
However, this spectral approach does not, in general, return the true maximum–weight clique defined above,
because it need not enforce the strict clique constraint (every pair of nodes connected). The underlying
optimization problem, MWCP, is NP–hard, and there are no known efficient algorithms for solving it exactly on
large-scale networks \cite{amgalan2014wmaxc}. As a result, traditional algorithms often incur substantial space
and time complexity. 
\nb{In addition to these general-purpose approaches, there exist specialised
maximum–weight clique heuristics and branch-and-bound methods tailored to
large combinatorial instances~\cite{cai2016fast}. Our goal here is not to
benchmark against the full ecosystem of such discrete algorithms, but to
demonstrate that the MWCP can be encoded and solved efficiently in a
gain-based analog platform and to validate this physical formulation
against representative digital baselines.
}

For instance, many classical methods rely on adjacency matrices to
accelerate adjacency checks, which becomes untenable for very large graphs; similarly, prevailing algorithms
typically do not exhibit the low time complexity necessary for rapid analysis at scale. In this light, it is
natural to seek physical solvers that can navigate the energy landscape of MWCP more efficiently, using the
intrinsic dynamics of analog systems to explore and select dominant subnetworks with reduced computational
overhead.

\begin{figure}
    \centering
    \includegraphics[scale=1]{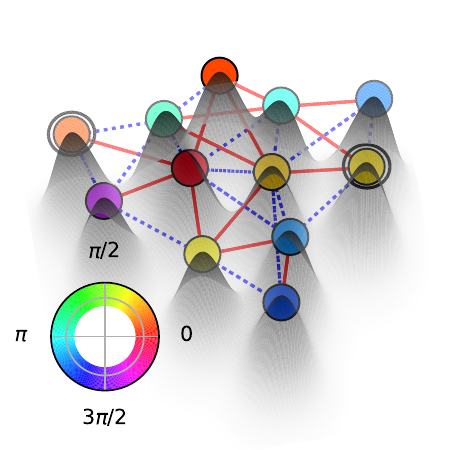}
    \caption{\ak{Schematic representation of the coupled condensates with fixed amplitudes}. The indirect positive and negative correlations obtained by minimizing the XY Hamiltonian using the temporal evolution of Eq.~(\ref{eq:laser-eq}) with Eq.~(\ref{eq:feedback}) and $\epsilon=0.1, \gamma_{\text{inj}}^{(i)}(0) = - 10$.
    Solid (red) edges between pairs of vertices $i$ and $j$ of the shown network represent ferromagnetic couplings  $J_{ij}=+1$, and the dashed (blue) edges correspond to anti-ferromagnetic couplings  $J_{ij}=-1$. Pairs of vertices not connected by edges have zero couplings $J_{ij}=0$. Colors indicate the phases of $\psi_i$ and reveal indirect correlations between the nodes. \ak{Thus, two circled nodes are highly correlated even though shortest route between them gives antiferromagnetic coupling ($J_{01} \cdot J_{12} \cdot J_{23}=-1$)}
}
    \label{fig:bio-xy}
\end{figure}

\nb{MWCP admits both a classical discrete formulation and an exact continuous encoding.
On a weighted undirected graph \(G=(V,E,w)\), the discrete maximum–weight clique
problem can be written as the integer quadratic programme
\begin{equation}
    \max_{y_i\in\{0,1\}} \;\sum_{i<j} w_{ij}\, y_i y_j,
    \quad 
    \text{s.t.}\; y_i y_j = 0\ \text{if} \quad (i,j)\notin E,
    \label{eq:IQP-MWCP}
\end{equation}
where \(y_i=1\) indicates that vertex \(i\) belongs to the clique.  This non-convex
binary optimisation is NP--hard and forms the standard combinatorial formulation of
MWCP.
The continuous formulation used in this work arises not from an ad-hoc relaxation,
but from the \emph{exact} Motzkin--Straus mapping between the discrete clique
problem and a quadratic programme over the simplex \cite{motzkin1965maxima}.  Under
this mapping, the binary variables \(y_i\) are replaced by continuous variables
\(x_i\ge 0\) subject to \(\sum_i x_i = 1\), yielding the non-convex quadratic programme
\begin{equation}
    \max_{x_i\ge 0} \sum_{i}\sum_{j<i} |J_{ij}|\, x_i x_j,
    \qquad \text{subject to }\sum_i x_i = 1.
    \label{eq:motzkin}
\end{equation}
Here \(N\) is the network size, \(x_i\in\mathbb{R}\) represents the continuous
importance of node \(i\), and the simplex constraint enforces normalisation of the
candidate subnetwork.  In the unweighted case \((|J_{ij}| = 1)\), the global maximiser
is exactly known: \(x_i = 1/\eta\) for all \(i\) belonging to a maximum clique of size
\(\eta\), and \(x_i=0\) otherwise \cite{motzkin1965maxima}.  For weighted networks,
all \(x_i\) outside the sought MWC vanish, while the nonzero \(x_i\) (with
\(\sum_i x_i = 1\)) encode the relative importance of nodes within the
maximum--weight clique.  Thus, the Motzkin--Straus formulation provides an
\emph{exact} continuous quadratic representation of MWCP, with no additional
relaxations beyond the original theorem.}

\section{Optical DOMINO platform}

Now, we formulate the "Optical DOMInant subNetwork detection" platform "DOMINO" and demonstrate its application to real biological gene-gene coexpression networks. In Eq.~(\ref{eq:motzkin}), we take the modulus of the coupling strengths $J_{ij}$ as it corresponds to the strength of the interactions (positive or negative). After identifying the dominant subnetwork, we show that the same optical platform can detect indirect correlations \ak{ expressed through phase differences} and quantify the strength and direction of these correlations between nodes. \ak{Figure \ref{fig:bio-xy} presents a schematic representation of the coupled condensates with fixed amplitudes, where different phases that minimize $XY$ model are illustrated using colored circles.}

To construct the optical GBC platform that solves MWCP of  Eq.~(\ref{eq:motzkin}), we consider the network of coupled condensates or lasers, a practical realization of PNNs,  with the state of the $i-$th condensate described by the wavefunction $\psi_i(t)=|\psi_i|\exp[i \theta_i]$ with phase $\theta_i$ relative to a reference condensate in a fast reservoir regime as
\begin{equation}
    \frac{d \psi_i}{d t} = \gamma_i(t,|\psi_i|^2) \psi_i + \sum_{j=1}^N J_{ij} \psi_j.
    \label{eq:laser-eq}
\end{equation}
This equation can be obtained as a tight binding approximation of the Ginzburg-Landau model \cite{kalinin2019polaritonic} with the effective gain (injection minus linear and nonlinear losses) given by $\gamma_i(t,|\psi_i|^2)$ and the interactions between condensates by $\mathbf{J}$.  The effective gain typically involves some form of a saturable nonlinearity, the simplest of which is $\gamma_i = \gamma_{\rm inj}^{(i)} - \gamma_{\rm loss} - |\psi_i|^2,$ with $\gamma_{\rm inj}^{(i)}>0$ and $ \gamma_{\rm loss}>0$ being the injection rate and linear losses, respectively. Various types of feedback can be envisioned, in particular, the injection rate that itself depends on the condensate density \begin{equation}
 \frac{d\gamma_{\rm inj}^{(i)}}{dt} = \epsilon(1 - |\psi_i|^2), \label{eq:feedback}   
\end{equation} which for some positive parameter $\epsilon$ brings all condensate densities to the same value ($|\psi_i|^2=1$). With such feedback, the steady-state of  Eq.~(\ref{eq:laser-eq}) was shown to minimize the loss function $F_0=\frac{1}{2}\sum_{i=1}^N(\gamma_{\rm inj}^{(i)} - \gamma_{\rm loss} - |\psi_i|^2)^2- \frac{1}{2}\sum_{i, j} J_{ij} \left( \psi_i \psi_j^* + c.c. \right) 
$ and, therefore, the XY Hamiltonian $H_{XY}=- \frac{1}{2}\sum_{i, j} J_{ij} \cos(\theta_i-\theta_j)$ \cite{kalinin2018networks}.

To solve MWCP 
 Eq.~(\ref{eq:motzkin}) we propose to associate $x_i$ with the amplitudes of $\psi_i$, assume that all condensates are phase-locked to the phase $\Phi$ of the reference condensate (so $\theta_i \rightarrow \theta_i-\Phi$) and 
 consider the optical loss function 
\begin{equation}
    F_{\rm loss} = \frac{1}{2}\sum_i\left( \xi - |\sum_j \psi_j|^2 \right)^2 - \frac{1}{2}\sum_{i, j} |J_{ij}| \left( \psi_i \psi_j^* + c.c. \right),
    \label{eq:complex-motzkin}
\end{equation}
where we modified $F_0$ by replacing  the local $|\psi_i|^2$ term in the effective gain  with the total light intensity $S=|\sum_j \psi_j|^2.$ \nb{ In this phase-locked regime we identify the continuous MWCP variables as
\begin{equation}
    x_i = \frac{\sqrt{\rho_i}}{\sum_j \sqrt{\rho_j}},
    \label{eq:x-rho-mapping}
\end{equation}
so that the penalty term in Eq.~(\ref{eq:complex-motzkin}) enforces the normalisation
\(\sum_i x_i = 1\) required by Eq.~(\ref{eq:motzkin}).
}

\ak{We express each complex oscillator amplitude $\psi_i$ in polar coordinates as $\psi_i=\sqrt{\rho_i} \exp \left(i \theta_i\right)$, where $\rho_i$ is the amplitude (intensity) and $\theta_i$ is the phase. Assuming all condensates are phase-locked (i.e., all $\theta_i$ equal), the optimization becomes a function of only the amplitudes, giving:}
%
%
$F_{\rm loss}=- \sum\limits_{i < j} |J_{ij}| \sqrt{\rho_i \rho_j} + \frac{1}{2} \left( \xi - |\sum\limits_j \sqrt{\rho_j}|^2 \right)^2$. \nb{With the identification in Eq.~\eqref{eq:x-rho-mapping}, the global minimum of
\(F_{\mathrm{loss}}\) coincides with the solution of Eq.~(\ref{eq:motzkin}), and the optical loss
functional faithfully encodes MWCP in the condensate amplitudes.} 
The system evolution  to the minimum of $F_{\rm loss}$ follows $\dot{\psi_i}=-\partial F_{\rm loss}/\partial \psi_i^*$, so that 
\begin{equation}
    \dot{\psi_i} =  (\xi - \;|\sum_j \psi_j|^2) \psi_i + \sum_{i\ne j} Q_{ij} \psi_j.
    \label{eq:physical}
\end{equation}
Equation \ref{eq:physical} describes the gradient–descent evolution of the coupled-condensate network towards the minimum of $F_{\rm loss}$, with effective couplings $Q_{ij} = |J_{ij}| - |\sum_k\psi_k|^2 + \xi.$ \nb{ The static physical couplings implemented in hardware are given by $|J_{ij}|$, while
\[
Q_{ij} = |J_{ij}| - \Bigl|\sum_k \psi_k\Bigr|^2 + \xi
\]
plays the role of an \emph{effective} dynamical coupling that combines the original edge weight with a global inhibitory contribution proportional to the total intensity.}

In experiments, the losses proportional to the total light intensity $S = |\sum_j \psi_j|^2$ can be achieved either through the absorption of the excited states of the condensates by the intracavity layer, resulting in correct energy blueshift \cite{blueshift}  or by measuring $S$  and adjusting the laser intensity using a feedback scheme on the effective gain  $\gamma_i(t)$.
Experimentally, the injected light must stay coherent with the reference condensate to satisfy the condition $\sum_i \psi_i = e^{i\Phi} \sum_i \sqrt{\rho_i}$. This can be achieved by applying an external coherent optical field \cite{schneider2013electrically} or implementing a feedback control system that monitors the phase of each condensate \cite{topfer2019towards}. Mathematically, we represent such coherence by adjusting 
 the phases of $\psi_i(t)$ in Eq.~(\ref{eq:physical}) at each evolution step by  the following feedback scheme (DOMINO)
\begin{equation}
    \small{\psi_i^{\rm new} = \left| \psi_i^{\rm old} + \Delta t\left( ( \xi - |\sum_j \psi_j^{\rm old}|^2) \psi_i^{\rm old} + \sum_{i<j} Q_{ij} \psi_j^{\rm old} \right) \right|}
    \label{eq:difference_scheme},
\end{equation}
%
where $\psi_j^{\rm old}$ and $\psi_j^{\rm new}$ are separated  by feedback time $\Delta t$.  The gain-based schedule arising in Eq.~(\ref{eq:difference_scheme}) modifies the basins of attraction and allows the trajectories to depart from the direction of the primary eigenvector and directs the trajectory to the true global minimum as illustrated in Fig.~\ref{fig:linear_penalty_evolution}.
Fig.~\ref{fig:linear_penalty_evolution}(a) shows the time evolution of amplitudes $|\psi_i|$ governed by Eq.~(\ref{eq:difference_scheme}). The insets in Fig.~\ref{fig:linear_penalty_evolution} demonstrate snapshots of the coupled condensate system at various times. 
Figure \ref{fig:linear_penalty_evolution}(b) shows histograms of solutions found by computer simulations of  DOMINO (Eq.~(\ref{eq:difference_scheme})) and its comparison with the 'BFGS' algorithm implemented via built-in 'scipy. optimize' library, as well as with solutions provided by the leading eigenvectors of the weights matrix $|J_{ij}|.$  To generate structured benchmark graphs with planted subnetworks, 
we first partition the vertex set into blocks of sizes 
$n_i \in \{3,4,5,6\}$ and construct fully connected cliques within each block. 
Random inter–clique edges are then added independently with probability $p$ to 
achieve the desired overall network density. Edge weights $J_{ij}$ are drawn 
uniformly from $[-1,0)\cup(0,1]$, and only connected graphs are retained.

\nb{One of the most critical advantages of DOMINO is
the efficiency with which it navigates the basins of attraction
towards the global minimum. This can be illustrated
by comparing the number of different random initial
conditions required to find the optimal solution using
DOMINO and a BFGS baseline.
For this comparison, we apply the BFGS quasi--Newton
method to the penalised Motzkin--Straus objective
\begin{equation}
    F_{\mathrm{BFGS}}(x)
    = -\sum_{i<j} |J_{ij}|\, x_i x_j
      + \lambda \Bigl( 1 - \sum_i |x_i| \Bigr)^2,
    \label{eq:FBFGS}
\end{equation}
with \(\lambda = 10\).
Here the \(x_i\) are continuous variables and the penalty
term enforces the simplex constraint \(\sum_i x_i = 1,\, x_i \ge 0\)
derived from Eq.~(1); in practice we observe that the
converged BFGS solutions remain close to the constraint
surface \(\sum_i |x_i| = 1\) for all tested instances.
DOMINO significantly exceeds the success probability of
the BFGS baseline, 
 as shown in Fig.~\ref{fig:linear_penalty_evolution}(c).}
 Figure \ref{fig:domino_gurobi_cliques_time} shows the accumulated probability of success depending on the execution time for DOMINO and a state-of-the-art quadratic programming solver Gurobi \cite{gurobi} applied to the original formulation of Eq.~(\ref{eq:motzkin}). It can be seen that DOMINO finds a solution with better confidence faster than Gurobi for structured complex networks with a density $p > 0.05$. \nb{We now turn to randomly structured weighted networks,
analysed in Sec.~III.
}

\section{Benchmarking on model networks}
\subsection{Performance on random networks}

To assess the generic performance of DOMINO beyond structured graphs with planted cliques, we also benchmark it on random weighted networks. The coupling matrices $\mathbf{J}$ are generated as
\[
J_{ij} = \frac{\eta_{i<j}(p) + \eta_{i<j}^T(p)}{2},
\]
where $\eta_{i<j}(p)$ is a random upper triangular matrix with entries $\eta_{i<j} \in [-1,0) \cup (0,1]$ drawn uniformly with probability $p<1$ and $\eta_{i<j}=0$ with probability $1-p$. The parameter $p$ thus controls the density of the weighted undirected network: vertices $i$ and $j$ are connected by an edge of weight $\eta_{i<j}$ with probability $p$. Only connected realizations of $\mathbf{J}$ are retained.

\begin{figure}[!htb]
    \centering
    \includegraphics[scale=0.35]{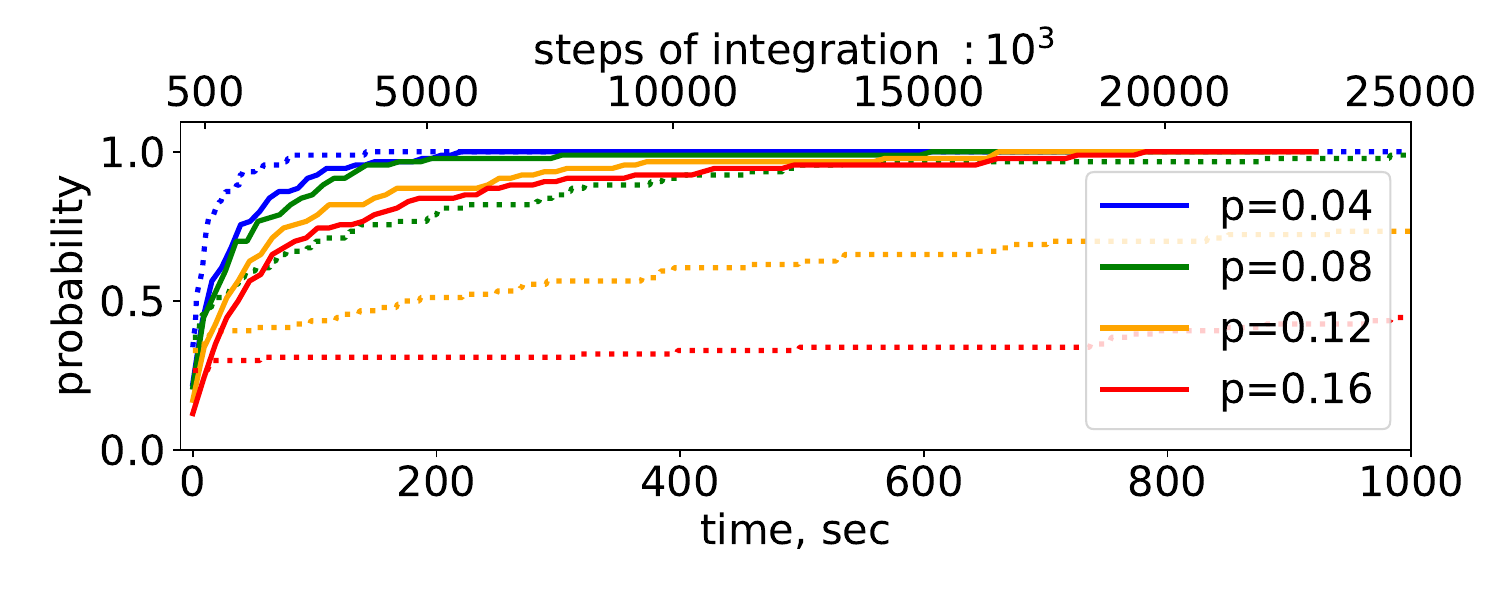}
    \caption{Accumulated probability of success as a function of time to solution (TTS) for Gurobi (dotted lines) and the DOMINO solver (solid lines) governed by Eq.~(\ref{eq:difference_scheme}) with $\xi=5$ and $dt=0.0002$. Both solvers are executed on the same single CPU core. For DOMINO, the TTS is estimated as the time required to sequentially execute the solver $100$ times, each starting from a different random initial condition. The upper axis shows the corresponding total number of integration steps of Eq.~(\ref{eq:difference_scheme}). The success probability at each time is defined as the fraction of $90$ different matrices $\mathbf{J}$ for which the global minimum is found.}
    \label{fig:domino_gurobi_time}
\end{figure}

Figure~\ref{fig:domino_gurobi_time} shows the accumulated probability of success as a function of the time to solution (TTS) for DOMINO and for the state-of-the-art commercial solver Gurobi, for networks of size $N=100$ and several values of the density $p$. For each network and each solver, the success probability at a given time is estimated as the fraction of $90$ independently generated matrices $\mathbf{J}$ for which the global minimum is found. Gurobi is run on a single CPU core, and DOMINO's TTS is estimated as the wall-clock time required to sequentially execute the solver $100$ times with different random initial conditions of Eq.~(\ref{eq:difference_scheme}).

\begin{figure}[!htb]
    \centering
    \includegraphics[scale=0.3]{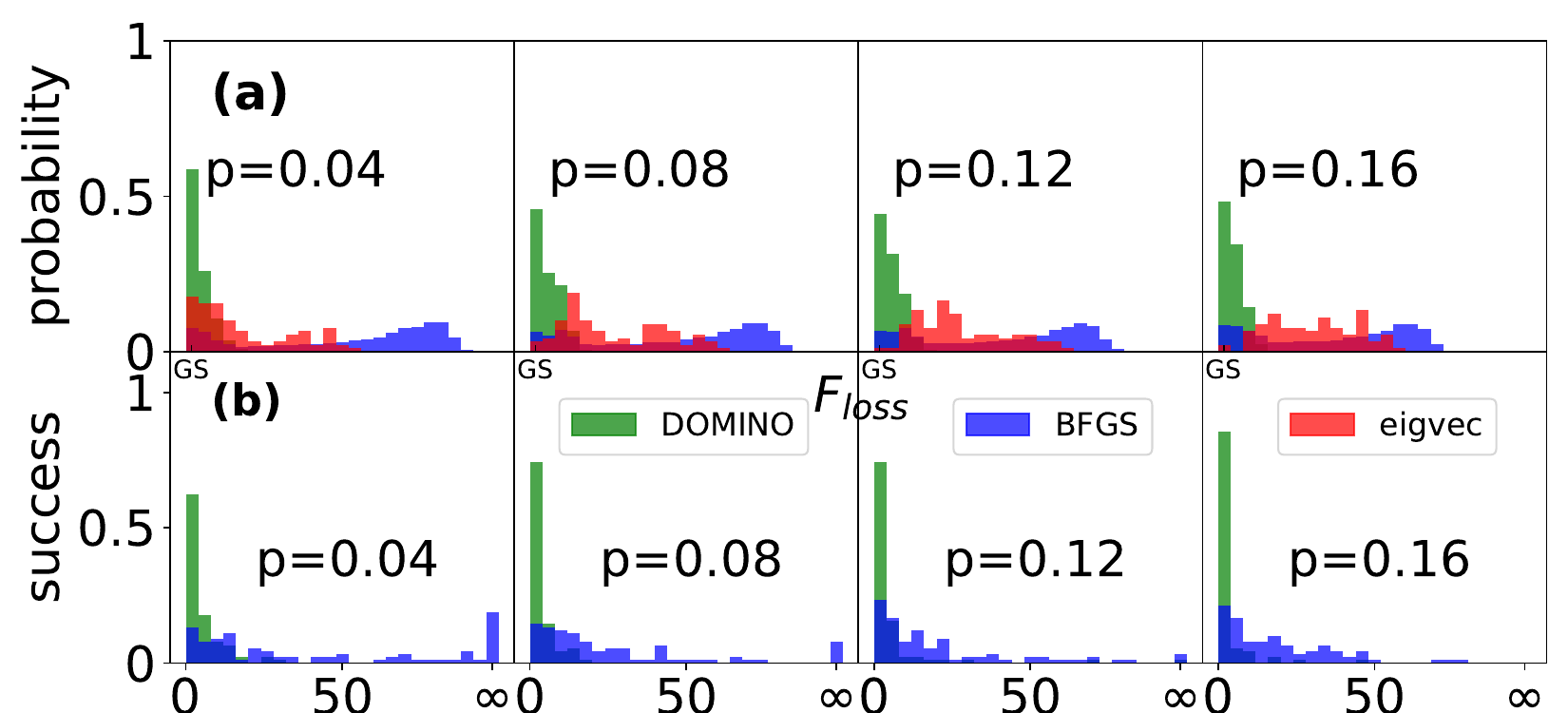}
    \caption{(a) Histograms of the objective values found by DOMINO (green) and by the BFGS algorithm (blue) for random networks of size $N=100$. For each of 100 independently generated matrices $\mathbf{J}$, 100 different random initial conditions are used. The red bars show the objective values obtained from the leading eigenvector of $\mathbf{J}$. The leftmost bars correspond to the global minima, verified using Gurobi. (b) Number of different random initial conditions required to find the global minimum at least once for 100 different connected random networks with $N=100$ and varying densities $p$. Green bars correspond to DOMINO (Eq.~(\ref{eq:difference_scheme})), and blue bars to BFGS. Values labeled $\infty$ indicate the fraction of networks for which the ground state is not found for any of the 100 initial conditions.}
    \label{fig:domino_gurobi_scipy}
\end{figure}

To compare DOMINO with gradient-based numerical optimization, Fig.~\ref{fig:domino_gurobi_scipy}(a) displays histograms of the objective values obtained by DOMINO, by the BFGS algorithm implemented in the \texttt{scipy.optimize} library, and by using the leading eigenvector of $\mathbf{J}$ as a heuristic proxy for the solution. The histograms are constructed from $100$ independent random networks of size $N=100$, with $100$ different random initial conditions per network. The leftmost bins correspond to the global minima, verified for each graph by Gurobi.

Figure~\ref{fig:domino_gurobi_scipy}(b) further quantifies the robustness of DOMINO with respect to initial conditions by showing, for each density $p$, how many different random initial conditions are required to find the global optimum at least once over $100$ trials for each of $100$ independent networks. 
\nb{These statistics can be interpreted directly in terms of the 
\emph{basins of attraction} of the underlying optimisation landscape.  
Because both DOMINO and BFGS follow gradient--based update rules on closely 
related continuous objectives, each random initialisation lies in the basin of 
attraction of some local minimum.  The empirical success probability therefore 
provides a measure of the \emph{relative volume} of the global basin.  The fact 
that DOMINO reaches the global optimum from far more initial conditions than 
BFGS indicates that its gain--controlled dynamics substantially enlarge the 
basin of the true solution.
}. In Fig.~\ref{fig:domino_gurobi_scipy}, bars marked by $\infty$ represent the fraction of networks for which the global minimum is not found within the $100$ sampled initial conditions. \nb{The fact that DOMINO
reaches the global optimum from far more initial conditions than BFGS
indicates that its gain--controlled dynamics substantially enlarge the
basin of the true solution.
Here BFGS is used not as a dedicated discrete clique solver, but as a
representative continuous optimization baseline probing the landscape
of the Motzkin--Straus objective in Eq.~(\ref{eq:motzkin}
In Fig.~\ref{fig:domino_gurobi_scipy}, bars marked by $\infty$ represent the fraction
of networks for which the global minimum is not found
within the 100 sampled initial conditions.
}

\begin{figure}[!h]
    \centering
    \includegraphics[scale=0.33]{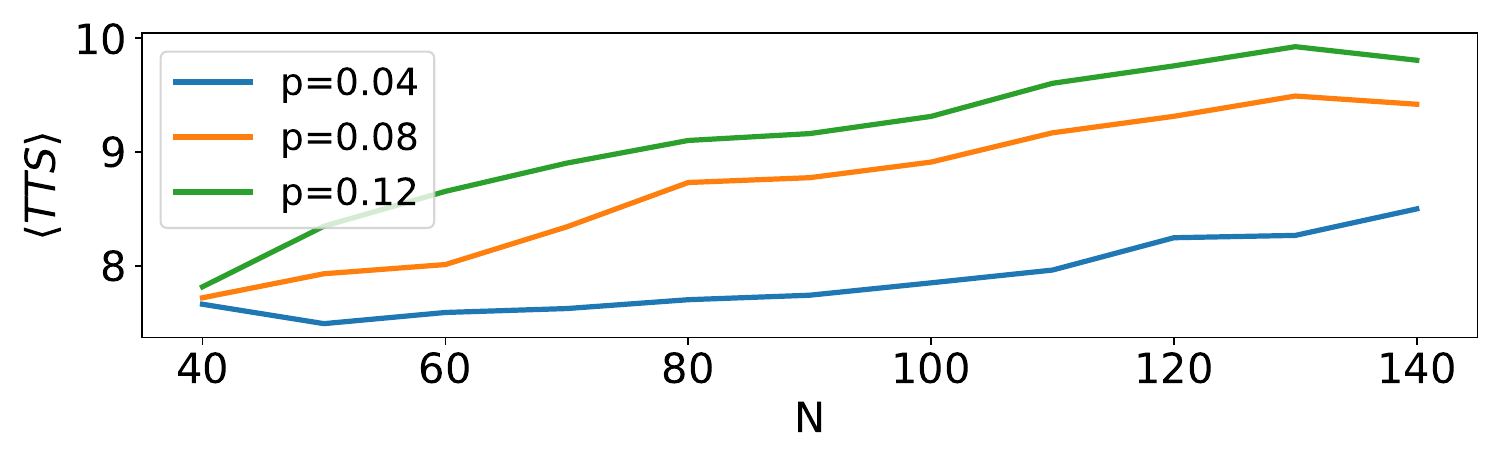}
    \caption{Average dependence of the time required for Eq.~(\ref{eq:difference_scheme}) with $\xi = 5$ and $dt = 0.0001$ to reach a steady state as a function of system size $N$. The average running time $\langle RT \rangle$ is obtained from 10 executions of DOMINO with different random initial conditions for each of 90 different matrices $\mathbf{J}$ at a given $N$ and density $p$. }
    \label{fig:domino_times}
\end{figure}

\begin{figure*}[t!]
    \centering
    \includegraphics[scale=0.55]{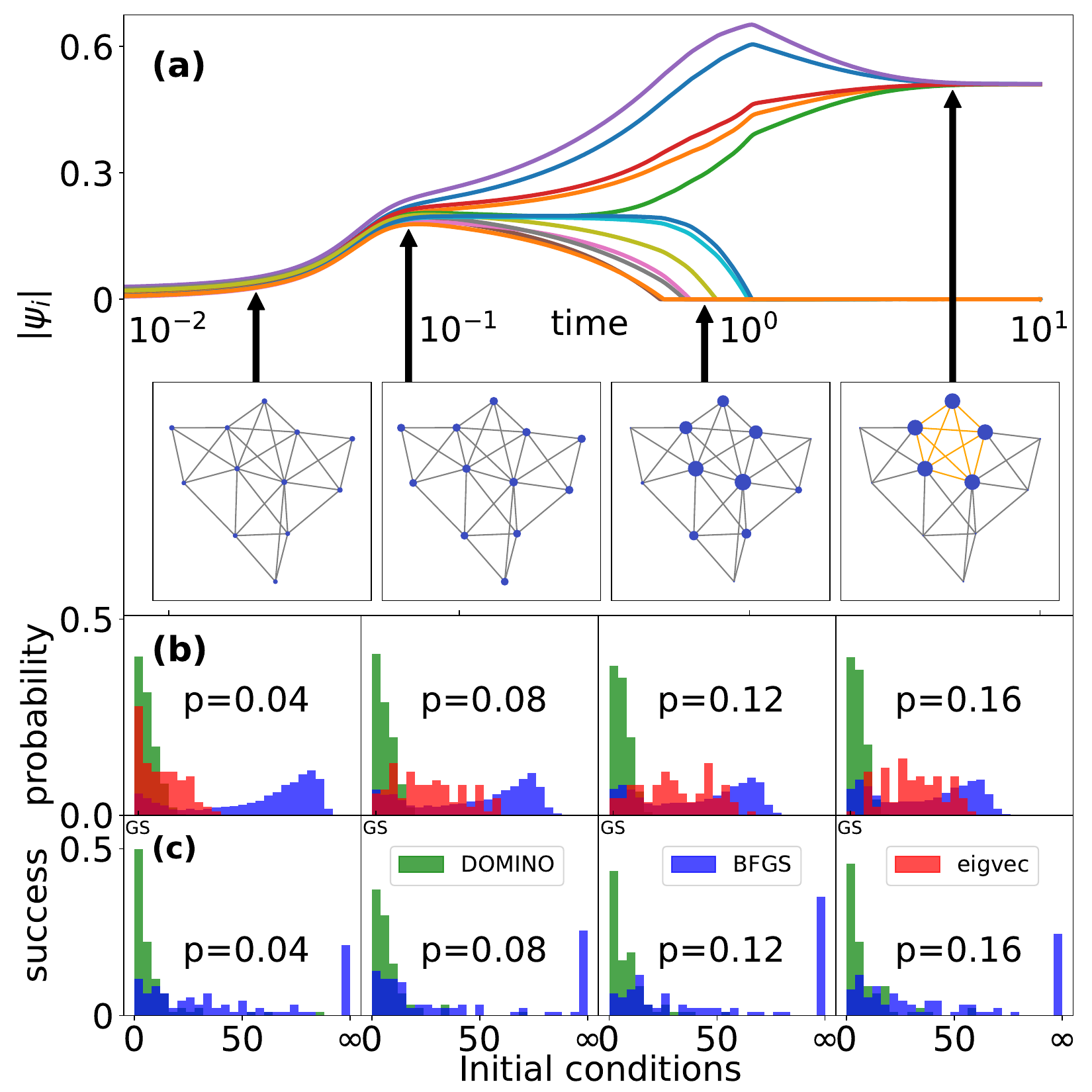}
    \caption{(a) Time evolution of amplitudes governed by Eq.~(\ref{eq:difference_scheme}) (with parameters $\xi=5$ and $dt = 0.0005$) for the network shown in the most left inset with $|J_{ij}|=1$ if there is an edge connecting $i-$th and $j-$th nodes and zero otherwise. 
    The insets show snapshots of the system of coupled condensates at various moments in time. Only the nodes in the maximum clique have nonzero amplitudes at the end of the simulation, as shown in the most right inset.
    (b): histograms of the solutions found by the trajectories of DOMINO  Eq.~(\ref{eq:difference_scheme}) (green) with parameters $\xi=5, dt=2 \cdot 10^{-4}$ and using 'BFGS' algorithm (blue). The red bars correspond to the solutions distribution provided by the leading eigenvectors of matrix $\mathbf J$.
    The histograms were calculated 
    based on $100$ connected random graphs of size $N=100$ with different weights matrices $J_{ij}$ generated for each of $100$ different initial conditions for each graph. 
    The leftmost bars correspond to the global minima, verified for each graph using the Gurobi solver. 
    (c): Number of different random initial conditions required to find the optimal solution using numerical integration of Eq.~(\ref{eq:difference_scheme}) calculated for $100$ different connected random networks of size $N=100$ with varying network densities $p$. Green bars correspond to solutions found by the DOMINO, and blue bars demonstrate convergence rates of the 'BFGS' algorithm. The values marked by $\infty$ show the fraction of networks $G$ for which the ground state was not found for any of the $100$ different initial conditions used.}
    \label{fig:linear_penalty_evolution}
\end{figure*}

The scaling of the characteristic time required by DOMINO to reach a steady state as a function of system size $N$ is shown in Fig.~\ref{fig:domino_times}. For each $N$ and density $p$, we average the running time $\langle RT \rangle$ over 10 executions of the solver with different random initial conditions for each of 90 independently generated matrices $\mathbf{J}$. The data indicate that the time to convergence grows with $N$ significantly more slowly than the quadratic growth in the number of edges, reflecting that the physical evolution of Eq.~(\ref{eq:difference_scheme}) does not explicitly perform $O(N^2)$ arithmetic operations.

\subsection{Robustness to noise}

Any physical implementation of DOMINO will inevitably operate with imperfect couplings due to fluctuations in pumping, finite accuracy of spatial light modulators, and overlap variations between condensates. To quantify the robustness of the solver to such imperfections, we introduce controlled noise into the coupling strengths and monitor the resulting degradation in performance.

We consider connected networks $G$ with coupling matrices $\mathbf{J}(G)$ generated as described above and solve Eq.~(\ref{eq:difference_scheme}) numerically with noisy couplings
\begin{equation}
J_{ij}^{\text{noisy}}(G) = J_{ij}(G) + \sigma \,\zeta_{ij}(t),
\end{equation}
where $\sigma$ sets the noise level and $\zeta_{ij}(t) \in (0,1]$ is a time-dependent random variable. If vertices $i$ and $j$ are not connected in $G$, we set $\zeta_{ij}(t)=0$, so that noise is only applied to existing edges.

To assess the impact of noise, we compare the loss function
\[
H_{\text{loss}} = -\frac{1}{2} \sum_{i,j} |J_{ij}| \left(\psi_i \psi_j^* + \text{c.c.}\right)
\]
for the noiseless system with
\[
H_{\text{loss}}^{\text{noisy}} = -\frac{1}{2} \sum_{i,j} \left|J_{ij}^{\text{noisy}}\right| \left(\psi_i \psi_j^* + \text{c.c.}\right)
\]
for the noisy system, evaluated at the corresponding steady states $\{\psi_i\}$. We define the relative error as
\[
\mathrm{Error} = \frac{H_{\text{loss}}^{\text{noisy}} - H_{\text{loss}}}{H_{\text{loss}}^{\text{noisy}} + H_{\text{loss}}}.
\]

\begin{figure}[!htb]
    \centering
    \includegraphics[scale=0.38]{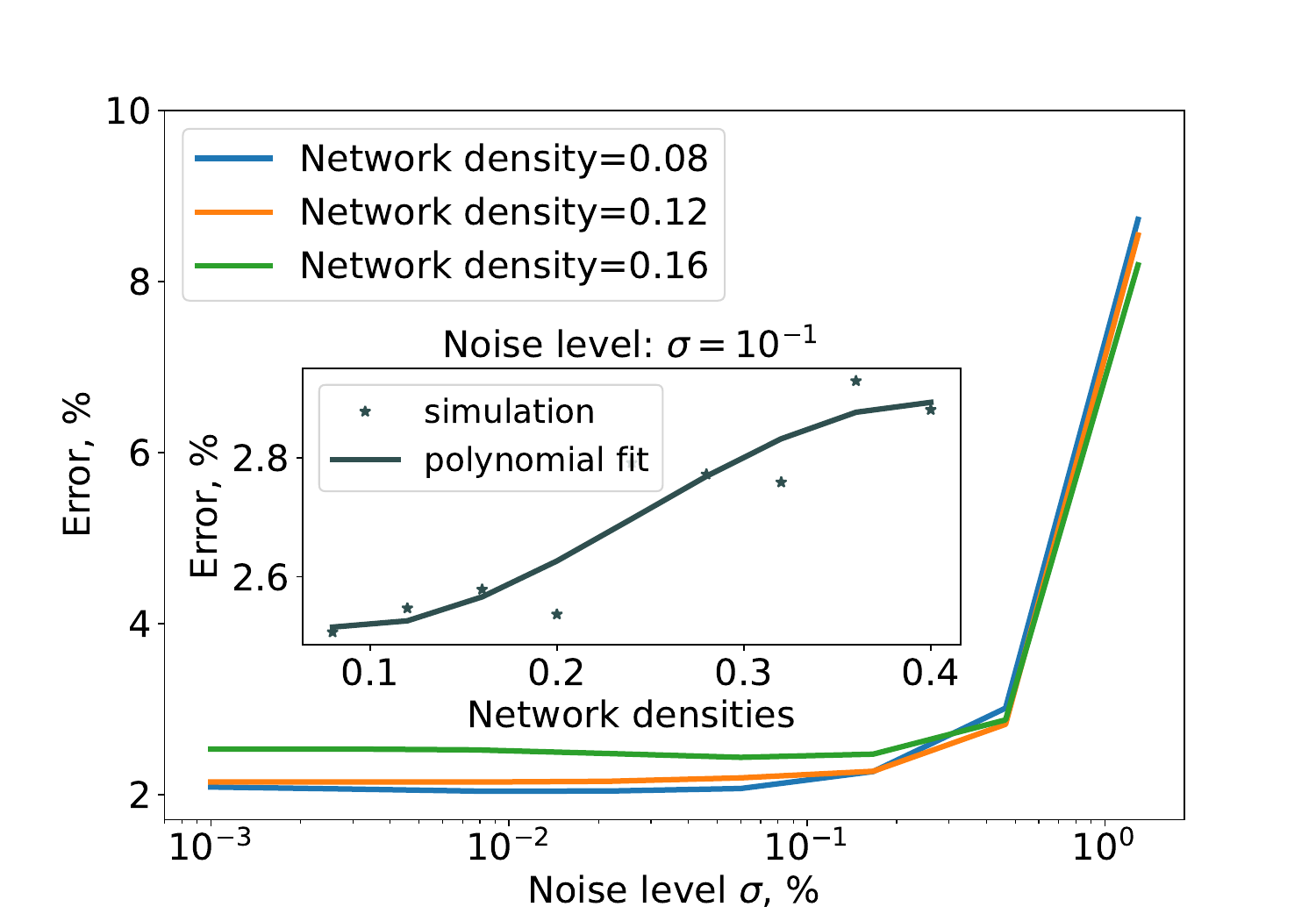}
    \caption{Robustness of DOMINO to noise in the couplings. Equation~(\ref{eq:difference_scheme} is solved numerically and averaged over $500$ different connected networks $G$ of size $N=30$  for several network densities $p$, with $100$ different random initial conditions per network. For each realization, the weights $\mathbf{J}(G)$ are perturbed by time-varying uniform noise of level $\sigma$ as $J_{ij}^{\text{noisy}}(G) = J_{ij}(G) + \sigma \zeta_{ij}(t)$, where $\zeta_{ij}(t)=0$ if vertices $i$ and $j$ are not connected in $G$ and $\zeta_{ij}(t) \in (0,1]$ otherwise. The plotted error quantifies the deviation of the loss function for the noisy system from the noiseless case, $\mathrm{Error} = (H_{\text{loss}}^{\text{noisy}} - H_{\text{loss}}) / (H_{\text{loss}}^{\text{noisy}} + H_{\text{loss}})$. DOMINO remains noise robust up to $\sigma \sim 1/2$. The inset shows the corresponding errors as a function of network density for fixed noise level $\sigma = 10^{-1}$.}
    \label{fig:robustness}
\end{figure}

\begin{figure}[!ht]
    \centering
    \includegraphics[scale=0.35]{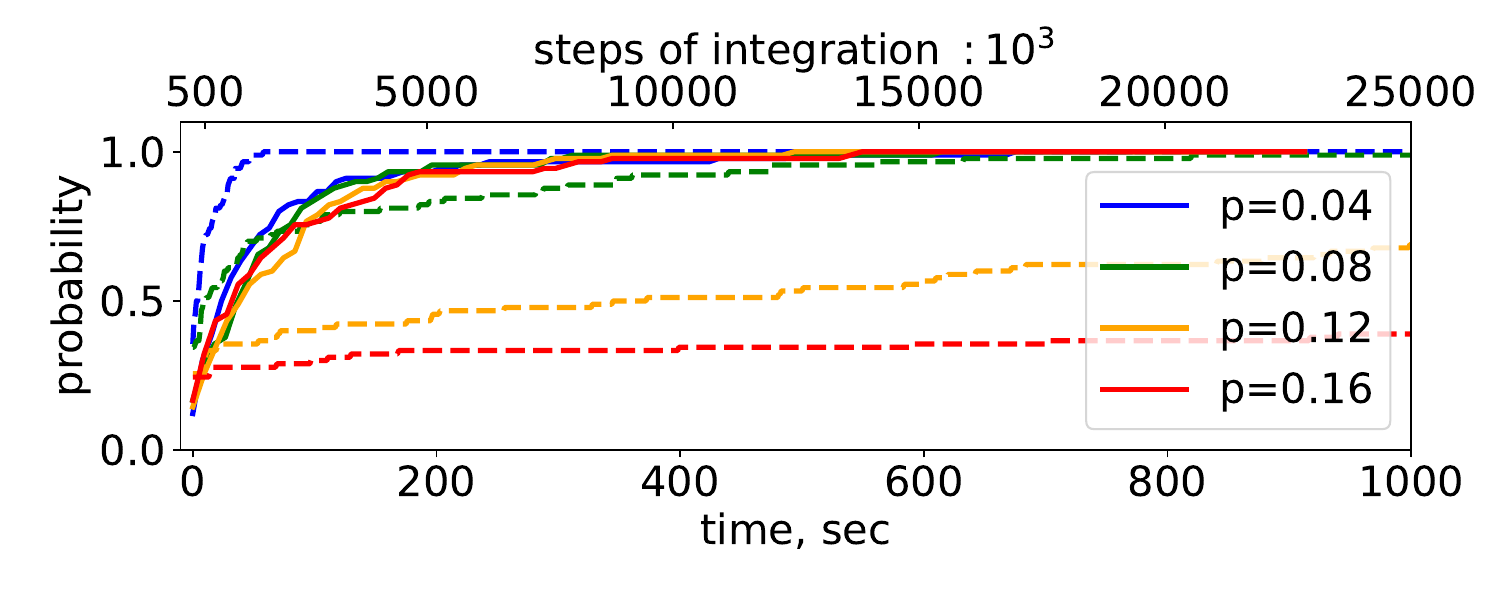}
    \caption{Comparison of the accumulated probability of success as a function of TTS for Gurobi (dashed lines) and DOMINO solver (Eq.~(\ref{eq:difference_scheme}) with  $\xi=5$ and $dt = 2 \cdot 10^{-4}$) (solid lines) for the networks of size $N=100$. Both solvers are executed on the same single CPU core. DOMINO computation time was estimated as the time required to sequentially execute solver $100$ times starting with different initial conditions. The upper axis shows the corresponding total number of steps of numerical integration of Eq.~(\ref{eq:difference_scheme}). The probability of success at each time step for both solvers was estimated as the fraction of $90$ different matrices $\mathbf{J}$ for which a global minimum was found. }
    \label{fig:domino_gurobi_cliques_time}
\end{figure}

Figure~\ref{fig:robustness} summarizes the results averaged over $500$ different connected networks $G$ of size $N=30$, with $100$ different random initial conditions per network. For each network, we perturb the couplings with time-varying uniform noise of level $\sigma$ and compute the corresponding error. We find that DOMINO remains robust up to noise levels $\sigma \sim 1/2$, with only modest deviations of the loss function from its noiseless value. The inset of Fig.~\ref{fig:robustness} shows the dependence of the error on network density $p$ for a fixed noise level $\sigma = 10^{-1}$.

\section{Application to gene coexpression networks}
\nb{The mapping between biological data and our physical solver proceeds as follows. 
First, we obtain gene expression levels from experimental data and, using these 
expression levels, construct a transcription–regulator (TR)–TR correlation 
matrix $W_{ij}$ (for example via the 3DCoop pipeline~\cite{yi2021interrogating}), which 
quantifies co-binding / co-expression relationships between TR pairs.  Second, 
we use this TR–TR matrix as the coupling matrix of the physical network by 
setting $J_{ij}=W_{ij}$, so that edge weights in the biological graph become 
coupling strengths between condensates.  Third, we encode the maximum–weight 
clique problem on this weighted TR network through the continuous Motzkin–Straus 
formulation in Eq.~(\ref{eq:motzkin}), with $|J_{ij}|$ as edge weights; in the 
optical implementation this objective is realised by the loss functional 
$F_{\text{loss}}$ and minimised by the dynamics of Eq.~(\ref{eq:difference_scheme}).  
Finally, once the system has relaxed to a steady state, the condensate 
amplitudes $|\psi_i|$ provide a direct readout of the dominant subnetwork: TRs 
with large steady–state amplitudes form the maximum–weight clique (the dominant 
coexpressed module), while subsequent evolution of the phases $\theta_i$ under 
the XY dynamics (Eqs.~(2)–(3)) reveals patterns of positive and negative 
(indirect) correlations between TRs.  Because the underlying optimisation 
landscape is non-convex, the physical solver may be initialised repeatedly from 
different random conditions to sample multiple basins of attraction and increase 
the probability of reaching the global maximum–weight clique.
} 

We applied our method to real-life biological data, specifically, a gene-gene coexpression network  \cite{yi2021interrogating}. This coexpression network was established using the 3DCoop pipeline, which analyzed the occupancy profiles of 386 Transcription Regulators (TRs) in the K562 human myelogenous leukaemia cell line \cite{K562}. The 3DCoop pipeline leveraged peak information obtained from ChIP-seq experiments, indicating the precise locations where the protein of interest binds to the DNA within the cell.
Utilising the information about how frequently these TRs bind to specific DNA locations within the cell as well as the spatial interactions between different genomic regions from Hi-C data, 3DCoop constructs TR-specific contact maps based on which the TR pairwise correlation matrix $W_{ij}$ of size $N\times N = 386 \times 386$ that was generated using the generalized Jaccard similarity. 
The solution of the MWCP can reveal which TR clusters  are involved in specific functions within the cell, including leukaemia or immune system-related functions (pathways). \nb{Conventional analyses of such data typically rely on methods such as weighted
gene-coexpression network analysis (WGCNA) and related module-detection
algorithms~\cite{langfelder2008wgcna}, which infer coexpression modules numerically
from transcriptomic profiles. In this work we instead focus on demonstrating that
DOMINO faithfully recovers the specific TR clique previously identified by
Yi~et~al.~\cite{yi2021interrogating} and, crucially, provides a direct physical route to solving the
underlying maximum-weight clique problem embodied in these correlation-based
networks.
}
The KEGG \cite{KEGG} pathways analysis can be used to interpret the cellular functions of identified TR cooperation in K562 cells (see Table S11 from \cite{yi2021interrogating}). \nb{The same mapping applies to any correlation-based gene network in which edge
weights quantify the strength of coexpression, co-binding, or other forms of
pairwise regulatory similarity between genes.
}

\begin{figure}[htb]
    \centering
    \includegraphics[scale=0.37]{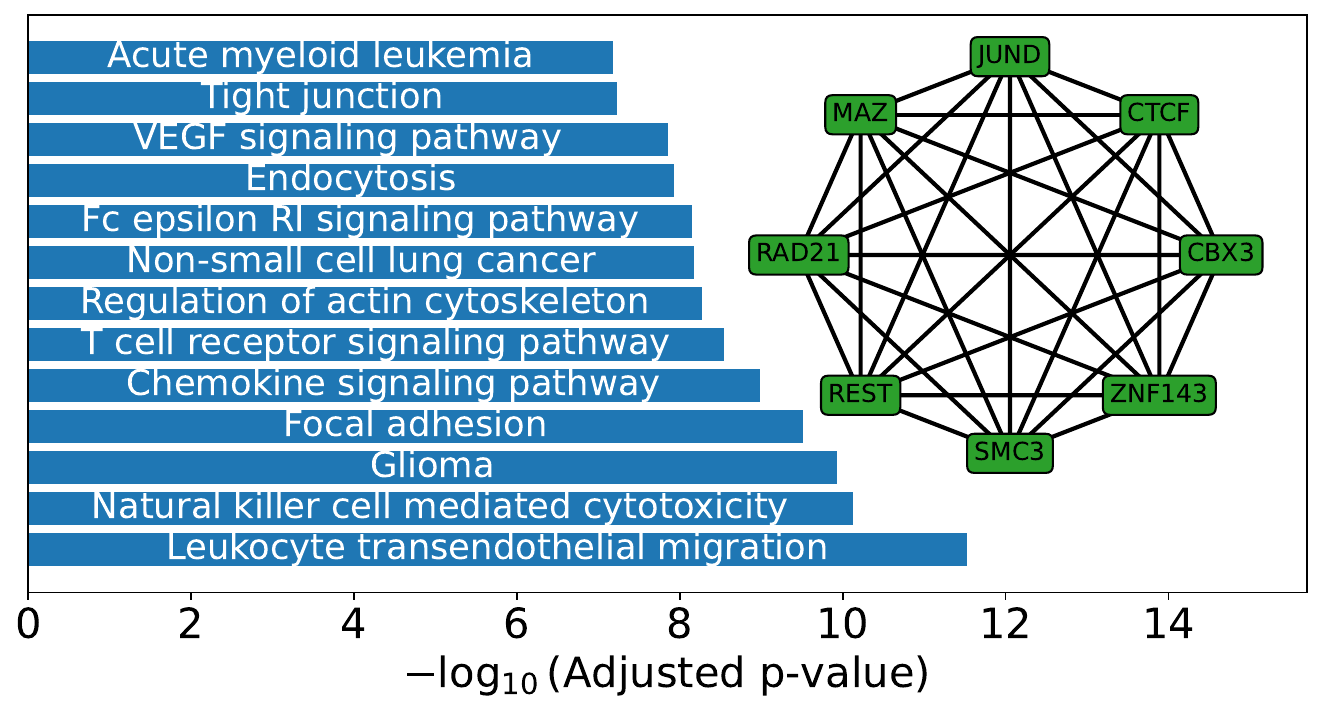}
    \caption{Seven most important KEGG pathways for ((CBX3-CTCF-JUND-MAZ-RAD21-REST-SMC3-ZNF143)) cluster identified as an MWC in the network of 386 Transcriptional Regulators detected in K562 human myelogenous leukemia cell line using 3DCoop pipeline. Inset illustrates the 
    most important Maximum Clique subnetwork found in the genes coexpression network of size $N=386$ from \cite{yi2021interrogating}. Green nodes 
    correspond to TRs highlighted by our solver. Calculations were performed with parameters: $\xi=5$, $T=20$, and $dt = 2 \cdot 10^{-4}$}
    \label{fig:enter-label}
\end{figure}

DOMINO was applied to identify the maximum weighted clique (MWC) based on the mean Jaccard index within the 3DCoop-generated network. The found maximum weighted clique is depicted in the inset of Fig.~\ref{fig:enter-label} and agrees with the findings of \cite{yi2021interrogating} (see Table S4 and Figure 2B).  
 Starting with $100$ different random initial conditions, our solver successfully found the most dominant clique in the gene-gene interaction network with $20\%$ success probability. At the same time, BFGS  had only $3\%$ success probability of solving Eq.~(\ref{eq:motzkin}) on the same set of initial conditions.
Figure \ref{fig:enter-label} shows the most enriched pathways associated with MWC found (CBX3-CTCF-JUND-MAZ-RAD21-REST-SMC3-ZNF143). This MWC agrees with Fig.~\ref{fig:domino_gurobi_time}(B) of \cite{yi2021interrogating}. The KEGG pathway analysis encompasses multiple pathways related to hematopoietic cancers. This finding implies that this collaboration of TRs may play a role in cell-type-specific gene regulation within leukaemia by influencing the 3D chromatin organization \cite{yi2021interrogating}. 

{\it Identifying and ranking all cliques} within a network based on their weights involves repeatedly applying MWC detection using DOMINO. After detecting an MWC, its vertices are iteratively deleted to form a new subnetwork, allowing the search for a new MWC to continue. This process is repeated for each clique found in the previous steps. To avoid redundant detections of the same clique, we keep track of the cliques already discovered and rank them according to their weights.

\section{Indirect correlations via phases}
DOMINO uses the condensate amplitudes to identify the most connected subsets in the network. In complex networks, direct connections may not exist between two nodes, but they can still exhibit correlations through other intermediary nodes and edges. \ak{This can point to shared regulatory pathways or responses to common stimuli}. \ak{This type of correlation is not captured by conventional co-expression analysis, which typically relies on direct pairwise interactions. Similarly, in sociological networks, phase alignment among unconnected individuals could indicate common influence or latent group affiliation, despite the absence of explicit links}. These indirect correlations  arise from various mechanisms like shared inputs/outputs, common regulatory factors, or feedback loops. Understanding and detecting these indirect correlations offer valuable insights into the network's structure and functionality. 
Once MWC is identified within a network, the phases of the oscillators in a complex-valued oscillatory network can be utilized to uncover both direct and anti- correlations among the network elements. To accomplish this, one can apply the original optical system, as outlined in Eq.~(\ref{eq:laser-eq}), which incorporates saturable nonlinearity, alongside the feedback mechanism described in Eq.~(\ref{eq:feedback}), designed to minimize the XY Hamiltonian. In this context, the coupling strengths between oscillators can be either positive or negative. The extent of indirect correlations between nodes is inferred from the magnitude of their phase differences as depicted in Fig.~\ref{fig:bio-xy}: a more significant phase disparity implies a stronger negative correlation. This method provides a systematic approach to evaluate the correlation strengths within the network. \nb{Applied to the transcription–regulator network analysed in Sec.~IV, such
phase patterns would, for example, highlight TRs that exhibit strong
synchronisation (small phase differences) despite weak direct
coexpression or connectivity in the underlying graph, suggesting shared
upstream regulators or participation in common pathways.  A systematic
phase-based characterisation of the K562 network, combining DOMINO’s
amplitude readout of maximum–weight cliques with the indirect correlation
structure encoded in the phases, is a natural direction for future work.
}

\section{Physical time and energy estimates}

Finally, we estimate the physical time and energy required by an optical implementation of DOMINO based on exciton--polariton condensates. The dynamics of a polariton condensate in dimensional units is well described by the complex Ginzburg--Landau equation
\begin{equation}
    i \hbar \frac{\partial \tilde{\Psi}}{\partial \tilde{t}} =
    - \frac{\hbar^2}{2 m} \tilde{\nabla}^2 \tilde{\Psi}
    + U_0 |\tilde{\Psi}|^2 \tilde{\Psi}
    + \frac{i \hbar}{2} \bigl( P_{\rm inj} - \gamma_c \bigr) \tilde{\Psi},
    \label{eq:Ginzburg-Landau}
\end{equation}
where $m$ is the polariton effective mass, $U_0$ is the strength of polariton--polariton interactions, $\gamma_c$ is the cavity loss rate, and $P_{\rm inj} = R_R P / (\gamma_R + R_R |\tilde{\Psi}|^2)$ describes the gain due to pumping, with $R_R$ the relaxation rate of reservoir excitations, $\gamma_R$ their loss rate, and $P$ the pump amplitude.

The dimensionless evolution equation (Eq.~(\ref{eq:laser-eq})) used in our simulations can be obtained via the scalings
\[
\tilde{\Psi} \to \sqrt{\frac{\hbar^2}{2 m U_0 l^2}}\, \Psi, \quad
\tilde{r} \to l r, \quad
\tilde{t} \to \frac{2 m l^2}{\hbar}\, t,
\]
followed by a tight-binding approximation for a network of coupled condensates. Within this mapping, a dimensionless evolution time $T$ corresponds to a physical time
\[
T_{\rm real} = \frac{2 m l^2}{\hbar}\, T.
\]
Using typical parameters for microcavity polaritons, $m \approx 10^{-4} \text{--} 10^{-5}\, m_e$ (with $m_e$ the free-electron mass) and a characteristic length $l = 1~\mu\mathrm{m}$, a dimensionless evolution time of $T = 100$ yields $T_{\rm real} \approx 270~\mathrm{ps}$ to $2.7~\mathrm{ns}$ per run for a single initial condition. As shown in Fig.~\ref{fig:domino_times}, $T$ grows with the network size $N$, but significantly more slowly than the number of edges $\sim O(N^2)$.

The energy cost of the purely optical DOMINO solver scales linearly with $N$ because the dominant resource is the creation and maintenance of $N$ overlapping condensates, each excited by a laser beam, and the interactions arise from their spatial overlap. A rough estimate for the optical energy consumption is
\[
E_{\rm real} \sim \text{const} \times \frac{\hbar^2 \gamma_{\rm inj} N}{2 m l^2} \propto O(N),
\]
where $\gamma_{\rm inj}$ is the effective injection rate. In contrast, electronic or optoelectronic optimisers typically consume energy per edge operation, leading to
\[
E_{\rm electronic} \approx E_{\rm el} \, p \, \frac{N(N-1)}{2} \sim O(N^2),
\]
with $E_{\rm el} \approx 1~\mathrm{pJ}$ a representative energy per logical operation and $p$ the network density~\cite{opticalNNs}. Exciton--polariton platforms thus offer an intrinsically energy-efficient route to solving maximum-weight clique problems, with both the physical time to solution and the energy consumption scaling favourably with problem size.

\nb{\section{ Related work and synergy}
A complementary direction for leveraging exciton--polariton platforms in graph problems was recently demonstrated in Ref.~\cite{wang2025polaritonic}, where lattices of condensates were used as \emph{feature–engineering front ends} for convolutional neural networks. In that approach, point clouds are embedded into pump landscapes, and the resulting photoluminescence and phase patterns---including vortex nucleation and interference fringes---serve as physically enriched positional embeddings that greatly improve topological–graph classification accuracy. Our DOMINO framework takes a fundamentally different but highly synergistic role: instead of preprocessing data for a digital classifier, DOMINO performs \emph{analog optimization}, directly solving the maximum weighted clique problem via the gain–controlled minimisation of the optical loss functional. The two paradigms are therefore complementary: DOMINO can supply physically generated clique solutions and dominant subnetworks as labels or inductive priors for large ML pipelines, while the photonic–ML embedding strategies of Ref.~\cite{wang2025polaritonic} can inform experimental designs for realising the coupling matrices $J_{ij}$ in optical hardware. This alignment highlights a broader emerging landscape, wherein polaritonic systems act both as fast analog optimisers and as physics–aware feature processors for graph-structured data.}

\section{Conclusion} 
\nb{We have developed DOMINO, a gain‑based light–matter computing platform that solves the maximum‑weight clique problem on weighted networks while simultaneously exposing indirect correlations through phase dynamics. Implemented as a network of coupled condensates, DOMINO combines a global‑intensity constraint with complex amplitudes to perform analog optimization directly in the continuous Motzkin–Straus landscape. On both structured and
random graphs, DOMINO achieves higher success probabilities and
shorter times-to-solution than a BFGS quasi--Newton baseline, and is
competitive with the state-of-the-art Gurobi solver, while remaining
robust to substantial noise in the couplings.
 Applied to a gene–gene coexpression network of 386 transcription regulators, the platform identifies biologically meaningful maximum‑weight cliques that are enriched in leukaemia‑related KEGG pathways, and reveals latent regulatory relationships encoded in the phase patterns. Together with favourable physical time and energy scaling, these results position DOMINO as a promising route towards energy‑efficient, hardware‑accelerated graph analytics for applications ranging from systems biology to finance and communication networks.}

\acknowledgements
A.K. thanks Cambridge Trust for support of his PhD studentship. N.G.B. is grateful to Prof 
Pooya Ronagh for a helpful discussion concerning mixed-integer programming using optics. N.G.B.~acknowledges the support from the HORIZON EIC-2022-PATHFINDERCHALLENGES-01 HEISINGBERG Project 101114978, the  EPSRC UK Multidisciplinary Centre for Neuromorphic Computing (grant UKRI982),   and Weizmann-UK Make Connection Grant 142568.

\bibliography{refsgene, references, ReferencesSpacialSpins,references_old}

\end{document}